\documentclass{aa}  
\usepackage[dvipsnames]{xcolor}
\usepackage{xspace}
\usepackage{graphicx}
\usepackage{txfonts}
\usepackage[pdftex,pagebackref,hyperindex=true,colorlinks=true,bookmarks=true,bookmarksopen=true,bookmarksnumbered=true,citecolor=blue]{hyperref}
\usepackage{siunitx}
\usepackage[normalem]{ulem}
\usepackage{float}
\usepackage{caption}
\usepackage{subcaption}
\usepackage{longtable}
\usepackage{xtab}
\usepackage{pdflscape}
\usepackage{adjustbox}
\usepackage{booktabs}
\newcommand{\OIII}{[O\,{\sc iii}]}
\newcommand{\NII}{[N\,{\sc ii}]}
\newcommand{\SII}{[S\,{\sc ii}]}

\newcommand{\HeI}{He\,{\sc i}}

\newcommand{\Ha}{H$\alpha$}
\newcommand{\Hb}{H$\beta$\ }
\newcommand{\kms}{\,\mbox{km}\,\mbox{s}^{-1}}

\begin{document}

   \title{Discovery and characterisation  of a new Galactic Planetary Nebula}


   \author{W. E. Celnik\inst{\ref{tbg}}
   \and 
   I. Karachentsev\inst{\ref{special}}
   \and 
   P. K\"{o}chling\inst{\ref{tbg}}
    \and
   S. Kotov\inst{\ref{special}}
   \and
   J. Kozok\inst{\ref{spectro}}
   \and
   L. Magrini\inst{\ref{arcetri}}
   \and
   A. Moiseev\inst{\ref{special}}
   \and
   M. Nischang\inst{\ref{tbg}}
    \and
   C. Reese\inst{\ref{tbg}}
    \and
   P. Remmel\inst{\ref{tbg}} 
    \and
   P. Riepe\inst{\ref{tbg}}
    \and
   T. Zilch\inst{\ref{tbg}}
          }

   \institute{TBG group of Vereinigung der Sternfreunde e.V., Postfach 1169, D-64629 Heppenheim (Germany)\label{tbg}
          \and
          Special Astrophysical Observatory, Russian Academy of Sciences, Nizhnii Arkhyz, 369167 (Russia)\label{special}
          \and 
          Spectroscopy group of Vereinigung der Sternfreunde e.V., Postfach 1169, D-64629 Heppenheim (Germany) \label{spectro}
          \and 
          INAF- Osservatorio Astrofisico di Arcetri, Largo E. Fermi, 5 50125 Firenze (Italy)\label{arcetri} \email{laura.magrini@inaf.it}
             }
\titlerunning{A new Galactic Planetary Nebula}
\authorrunning{Celnik et al.}

   \date{Received ; accepted }

 
  \abstract
   {Planetary nebulae are one of the final stages in the evolution of low and intermediate mass stars. They  occur in a variety of shapes. Older and fainter  ones are generally more difficult to identify because of the lower surface brightness.   } 
   {This paper reports the serendipitous discovery of a new faint Galactic planetary nebula (PN),  during a campaign to identify dwarf galaxies,  companions of the spiral galaxy NGC~2403. We aim at confirming the nature as PN of a diffuse object identified in the Camelopardalis constellation.  }
   {We obtained narrow-band filter images and spectra of the nebula and its central star with amateur and professional telescopes having  diameters from 20 cm to 6 m. }
   {We detected a dense triangular nebula, surrounded by an elliptical region, named Cam nebula. They are part of a larger and fainter circular nebular structure, named TBG-1, at the centre of which we have identified the possible central star, a white dwarf with a temperature of about 22\,000 K. The analysis of the spectrum made it possible to measure the physical characteristics of the nebula, in particular its electronic density and temperature.  }
   {Analysis of the images, of the spectra of the nebula  and of the central star confirm the PN nature of TBG-1, located at the distance of about 1 kpc. This work reaffirms the potential for fruitful collaborations between astronomers and amateur astronomers in the detection and study of new objects. }


   \maketitle
%

\section{Introduction}

Planetary nebulae (PNe) are the remnants of  stars with initial masses between $\sim$1 and 8~M$_{\odot}$, which expel their outer layers during the latest phases of their evolution. The PN phase is quite short in terms of astrophysical times \citep[e.g.][]{Badenes2015ApJ...804L..25B}, but since they are produced by the most representative stars of the initial mass function (IMF), PNe are quite common in our Galaxy \citep[e.g.][]{Acker1992secg.book.....A, Kerber2003A&A...408.1029K, Stanghellini2020ApJ...889...21S, Gonz2021A&A...656A..51G} and in external galaxies \citep[e.g.][]{magrini2000A&A...355..713M, magrini2003A&A...407...51M, magrini2005MNRAS.361..517M, pena2007A&A...466...75P, pena2012A&A...547A..78P, herna2009A&A...495..447H}.  
The number of PN detections has grown considerably in recent years \citep[see][for the MQ/AAO/Strasbourg Galactic PNe database MASPN
 and its updates]{Acker1992secg.book.....A,  catalog2006MNRAS.373...79P, catalogue2014apn6.confE..69P}, thanks also to the new  PNe discovered with wide-field observations in the narrow-band H$\alpha$  both in the  Northern and Southern Galactic plane \citep{catalogue2014MNRAS.440.2036D, Sabin2014MNRAS.443.3388S, Sabin2021MNRAS.508.1599S,  Sun2023MNRAS.tmp.3785S}.
More recently, a multi-band vision allowed us for an ever-widening knowledge of the PNe population, such as the UV studies with Galaxy Evolution Explorer (GALEX) UV sky surveys \citep{galex2019Ap&SS.364..181P, galex2023ApJS..266...34G} or with the Cornish radio catalogue \citep{cornish2018MNRAS.480.2916F}, with the discovery and characterisation of new objects.  
In addition, new artificial intelligence techniques are giving new impetus to the PN search, greatly increasing their numbers, up to 20000 possible candidate PNe \citep{Sun2023MNRAS.tmp.3785S}.

However, the detection of new PNe and symbiotic stars  is not only based on the work of professional astronomers, but also many amateur astronomers are actively contributing \citep[e.g.][]{acker2014LAstr.128a..40A, ledu2022A&A...666A.152L, petit2023NewA...9801943P, ritter2023arXiv231103700R, dre2023RNAAS...7....1D, fesen2023ApJ...957...82F}.


In this context, German amateur astronomy groups are also very active. The Tief Belichtete Galaxien (TBG) which means "Long exposed galaxies"\footnote{http://tbg.vdsastro.de/} group is part of the {\em VdS-Fachgruppe Astrofotografie} group, one of the 19 working groups in the German amateur association {\em Vereinigung der Sternfreunde}\footnote{https://sternfreunde.de/}. In 2012, the TBG group and the Astronomical Institute of Bochum Ruhr University established a collaboration, in order to search for new extragalactic objects in the Local Volume, such as stellar streams or new dwarf galaxy candidates. 
The collaboration proved to be fruitful providing the identification of  some new dwarf galaxy candidates \citep{Karachentsev15, Karachentsev20, Blauensteiner17}.

\begin{figure}
   \centering
   \includegraphics[width=\hsize]{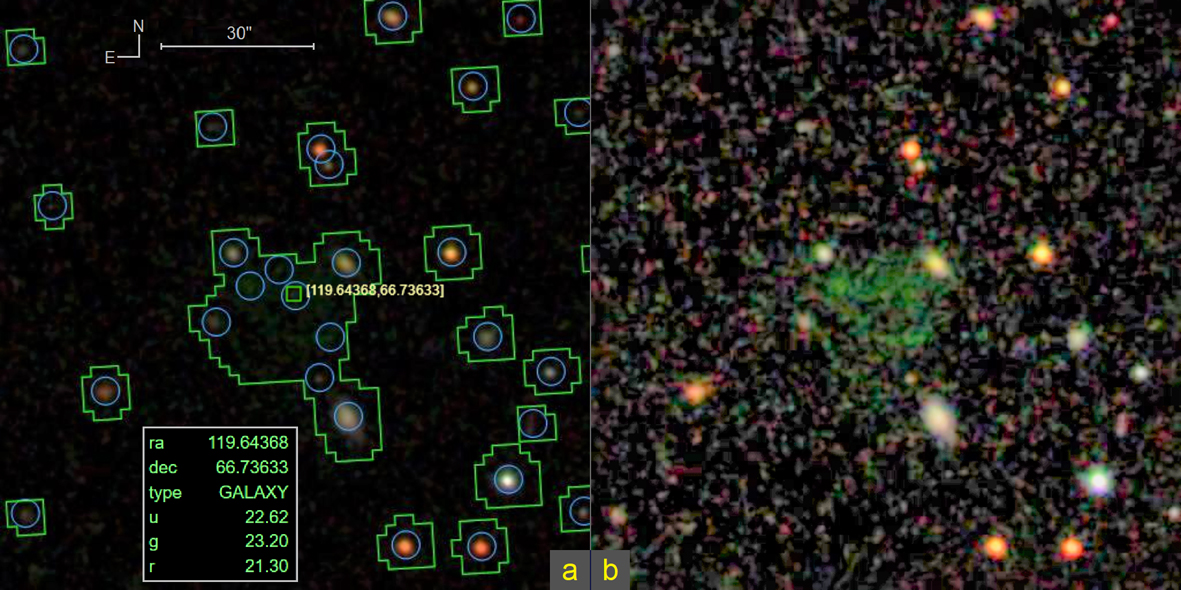}
      \caption{Field (100\arcsec$\times$100\arcsec) around RA = 119.6414$^{\circ}$ and Dec = 66.7361$^{\circ}$ in the original SDSS (DR16) frame. Left panel: the  SDSS view of the field, with objects with photometry  marked by light blue circles and SDSS outlines are green. According to the SDSS classification, the object in the centre of the field is a galaxy (see inset with SDSS information: ra, dec, type and photometric mag u, g and r). Right panel: The same SDSS field with the same image scale, brightened and contrast enhanced without any markings, for which a  green diffuse object of $\sim$30\arcsec\, is visible.
              }
         \label{fig:1}
   \end{figure}

From a visual inspection of the 16$^{th}$ data-release   of the Sloan Digital Sky Survey  \citep[SDSS DR16,][]{sloandr16} with the aim of 
identifying dwarf galaxy companions of the spiral galaxy NGC 2403, the TBG group detected at the position RA = 119.6414$^\circ$ (07 h 58 min 33.9 s) and DEC = 66.7361$^\circ$ ($+$66$^{\circ}$ 44\arcmin 10\arcsec) in the constellation of Camelopardalis a very faint object with a triangular shape (Fig.~\ref{fig:1}). 
As per automatic classification of SDSS (see the inset in Fig.~\ref{fig:1}), the irregular shape and the angular size of the object ($\sim$30\arcsec) could correspond to that of a small irregular galaxy (with a diameter of 480 pc), a possible satellite of the nearby spiral galaxy NGC~2403 at the distance of 3.3 Mpc  \citep[see., e.g.][]{Karachentsev04}. 
The green colour of the detected object (see the panel to the right in Fig.~\ref{fig:1}) corresponds to the [O~{\sc iii}] emission line, which is typical of H~{\sc ii} regions, usually present in star-forming irregular galaxies. 
Therefore, to investigate  the nature of this object, an observational campaign was carried out by the TBG group.  
In particular, we obtained new deep (H$\alpha$+[N~{\sc ii}]) observations  which highlighted the presence of the nebular object  and located it  within a fainter elliptical nebula extending 5\arcmin.7 from northeast to southwest direction (Fig.~\ref{fig:2}). The presence of this elliptical nebula, which is not detectable in the SDSS images, have weakened the initial hypothesis of an irregular galaxy.  The elliptical nebula is named from here onwards  Cam nebula.

In this paper, we present the discovery of a new candidate planetary nebula with an angular diameter of 9\arcmin  and an extremely faint circular shell,  named TBG-1. This shell includes the much brighter but smaller Cam nebula at TBG-1´s southern border. The identification and subsequent photometric and spectroscopic observations of this object provide information on the nature of its central star and of the surrounding gas. 
The paper is structured as it follows: in Section~\ref{sec:obs} we describe our observations, both imaging and spectroscopy; in Section~\ref{sec:red} we give details on the data reduction process, whereas in Section~\ref{sec:analysis} we describe our analysis. In Section~\ref{sec:centralstar} we describe the properties of the nebula and of the central star.  In Section~\ref{sec:conc} we present our summary and conclusions.


\section{Observations}
\label{sec:obs}

\subsection{Imaging}

We observed the nebula from February 2022 to June 2023 with the TBG group telescopes, having  diameters from 200~mm to 356~mm (apertures from f/2 to f/7.7). 
We obtained deep exposures with narrow-band filters centred at 656 nm  (H$\alpha$+[N~{\sc ii}]),  at 501 nm ([O~{\sc iii}]),  at 672 nm ([S~{\sc ii}]), at 546 nm (continuum near [O~{\sc iii}]), and at $\sim$880 nm in the near infrared (NIR, edge position of the long wave pass filter 850 nm). 
We stacked relatively short single exposures (between 60 and 1200 seconds, depending on the bandwidth and aperture) to obtain a final image per filter and per night.  In summary, we obtained  3245 individual exposures in the H$\alpha$+[N~{\sc ii}] filter in 25 nights of observations, 1606 exposures in the  [O~{\sc iii}] filter in 12 nights, and 348 exposures in the [S~{\sc ii}] filter in 3 nights.
We obtained  also 1284 individual exposures with a total of 24 hours exposure during 4 nights  in the NIR. 
Finally, photo images  were taken with mono cameras and luminance filters as well as with colour cameras without filters. 
The total exposure includes 8819 individual exposures in 61 nights with a total exposure of 277 hours.

In addition to the observations with TBG amateur telescopes, 
we observed  the brightest parts of the nebula  with narrow-band filters centred in  H$\alpha$  and in the continuum near to the H$\alpha$+[N~{\sc ii}] lines with the 6-m Big Telescope Alt-Azimuth (BTA) telescope of the Special Astrophysical Observatory of the Russian Academy of Sciences (SAO RAS) using SCORPIO-2 multimode focal reducer  \citep{Afanasiev2011} providing flux-calibrated image in $6.4\times6.4$ arcmin field of view with a sampling $0.4$ arcsec per px.

To complete the data reduction, we collected  several images every night and for each telescope/filter configuration: flat fields, darks for science  and for flat fields and bias images.

\subsection{Spectroscopy}
\begin{figure}
   \centerline{
    \includegraphics[height=\hsize]{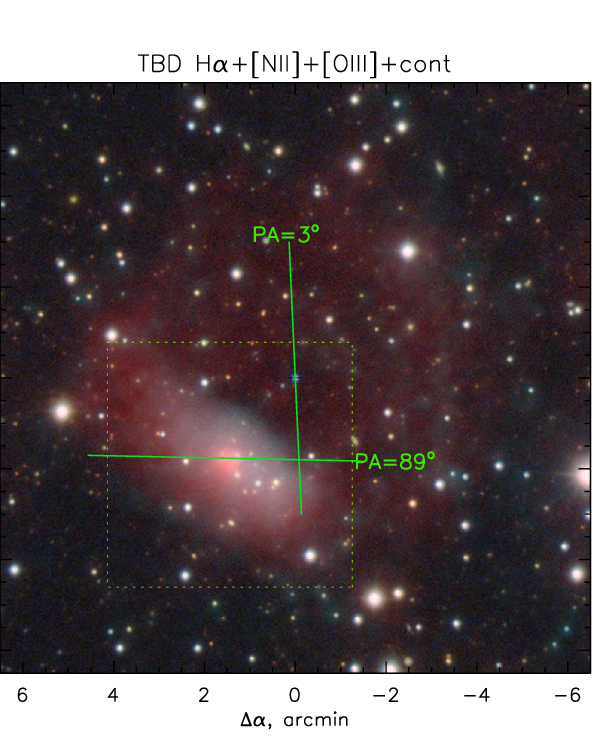}
   } 
      \caption{Combined TBG-1 image  with the positions of the two SCORPIO-2 slits with $PA=89\degr$ and $PA=3\degr$. North is up, east to the left.}
                   \label{fig:slits}
  \end{figure}

The long-slit spectra were obtained with the SAO RAS  6-m  telescope  BTA  using the  SCORPIO-2 multimode focal reducer  with the $6\arcmin\times1\arcsec$ slit providing the spectral resolution $\sim5$\AA\, in the spectral range 3650-7300 \AA. The spatial scale was $0.4$\arcsec/px. 
The spectrum across the brightest part of the nebula was obtained on   April 7$^{th}$, 2022  with a position angle $PA=89\degr$ (see Fig.~\ref{fig:slits}), a total exposure time $3000$ s and atmospheric seeing 2\farcs3. The spectrum of the central star  was obtained on   March 22$^{th}$, 2023  with the slit placed along atmosperic dispersion angle  ($PA=3\degr$), a total exposure time $1500$ s and  seeing 2\farcs0.

\section{Data reduction}
\label{sec:red}

\subsection{Imaging data reduction}
Each observer produced a science image per observation night and per filter: first, master flats, darks and biases were created and applied to each individual image before the science images were stacked. Then, the science frames were averaged using stacking programs,  
removing cosmic rays,  and producing a final image per night and per filter.

The 25  images in the H$\alpha$+\NII{} filter 
were stacked 
to produce a final image.  The same procedure was adopted for the images resulting from  the 3 nights in the [S~{\sc ii}] filter,  and for the 12 images in the  [O~{\sc iii}] filter with f/2 (the 6 stacked images with lower aperture were too faint to be combined). 
We obtained two final images, one  in the H$\alpha$+\NII{} filter with a total exposure time of 115 hours  and one in the [O~{\sc iii}] filter with a total exposure time of  64 hours.  The image in the [S~{\sc ii}] filter, with  12 hours of exposure, was too noisy to enable subsequent analysis.

\begin{figure}
   \centering
   \includegraphics[width=\hsize]{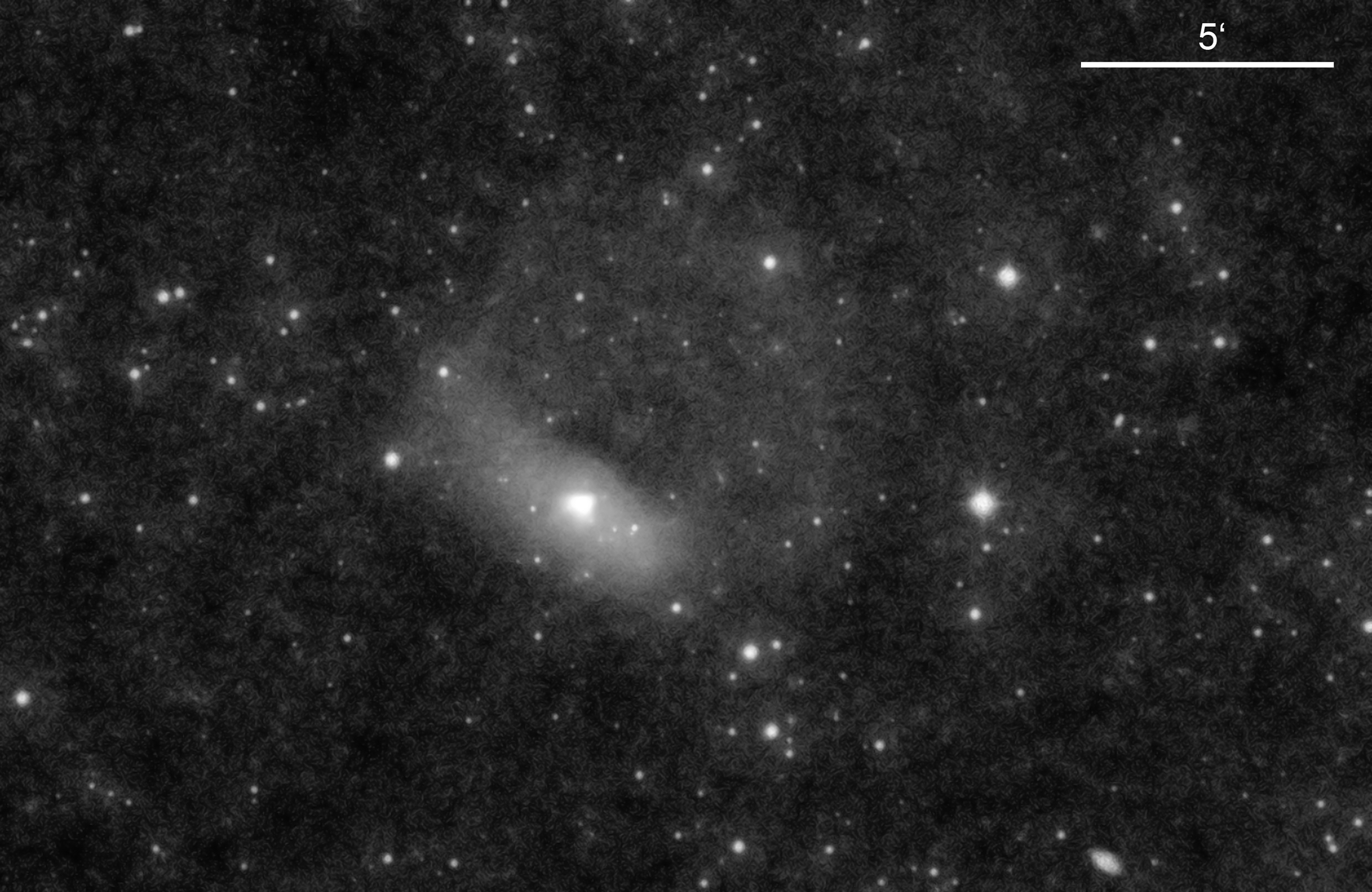}
      \caption{Combined image  in the H$\alpha$+\NII\, filter of the whole nebular complex TBG-1.    The total exposure time is 115~hr. North is up, east to the left.
              }
         \label{fig:2}
   \end{figure}

\begin{figure}
   \centering
   \includegraphics[width=\hsize]{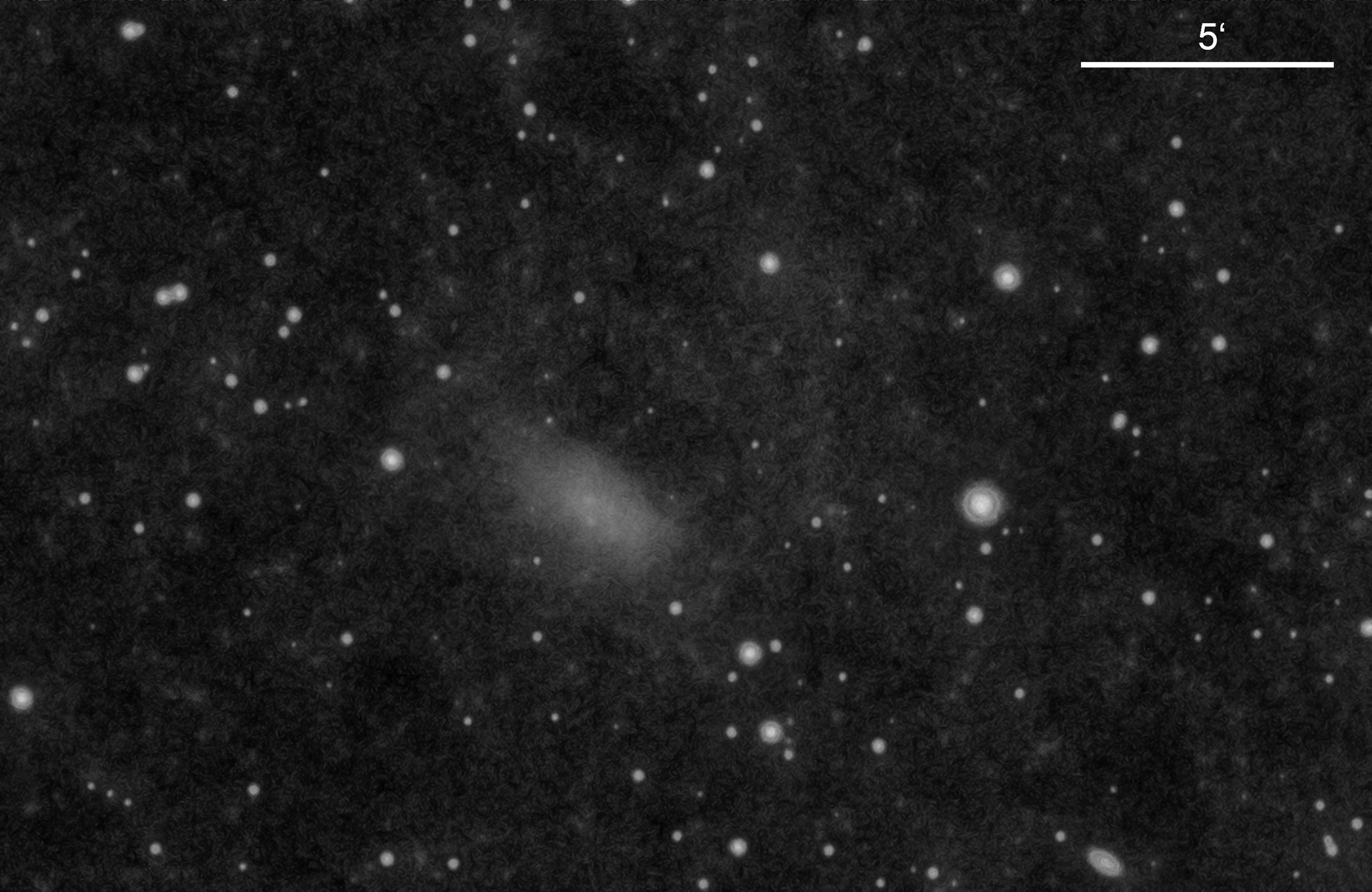}
      \caption{Combined image in the \OIII\, filter of the whole nebular complex TBG-1. The total exposure time is 64~hr.  North is up, east to the left.
              }
         \label{fig:3}
   \end{figure}

\begin{figure}
   \centering
   \includegraphics[width=\hsize]{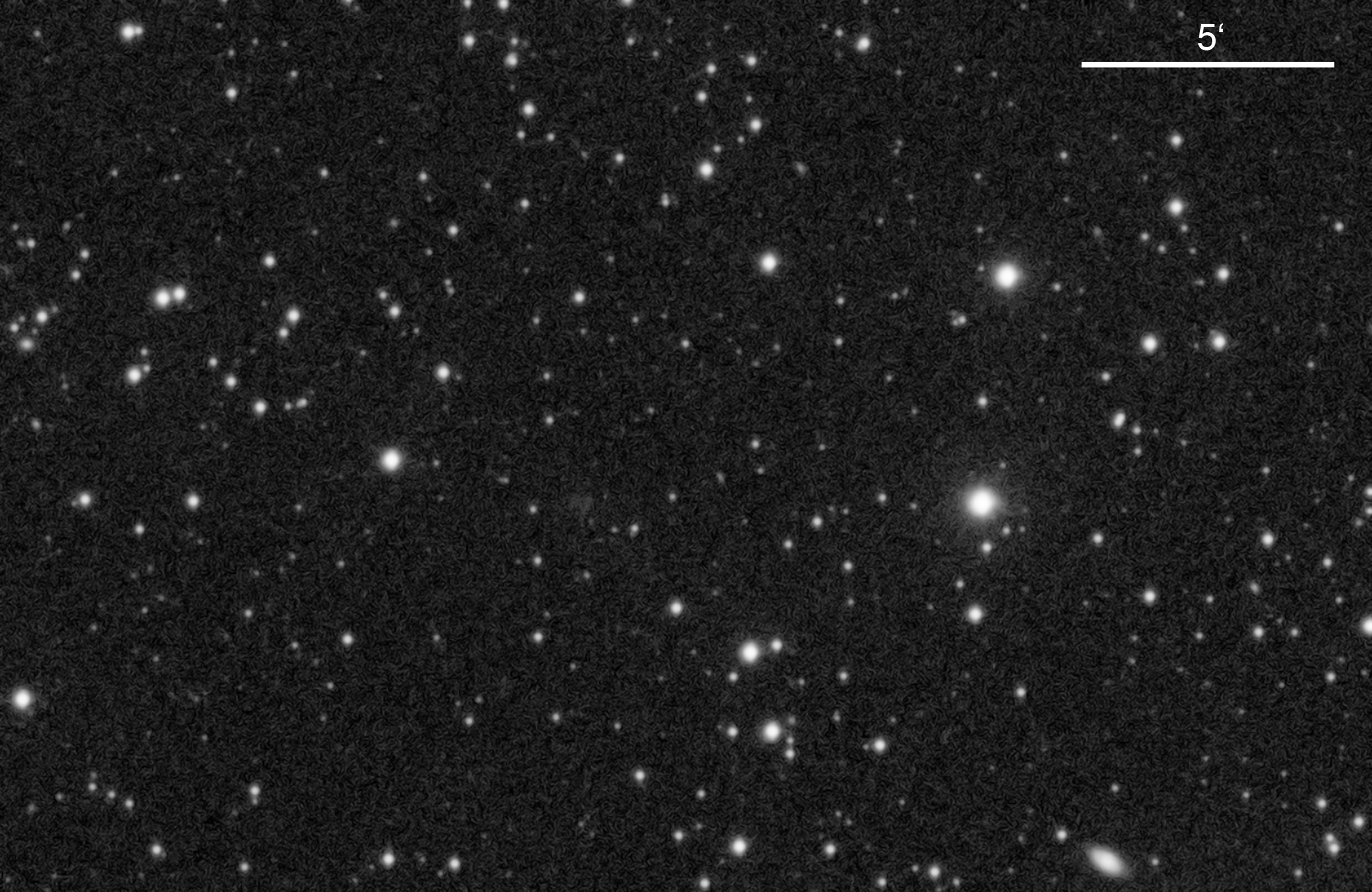}
  \caption{Combined image in the  \SII\, filter of the whole nebular complex TBG-1. The total exposure time isr 12~hr.  North is up, east to the left. }
         \label{fig:4}
   \end{figure}

Since the emission of the nebula is extremely weak, it is not detectable with a linear brightness scale. We therefore used a photographic technique that consists of combining images on a logarithmic scale to maximise contrast and show better the nebular features. To do that,  we  aligned and cropped with {\sc Photoshop} \citep{adobephotoshop} the  25  images in the H$\alpha$+\NII{} filter,  using a non-linear brightness scale. We produced a mask with the location of the stars using {\sc Astro Pixel Processor}\footnote{https://www.astropixelprocessor.com/}, and we applied it to the images, obtaining a starless image per night. 
The starless individual images showed only the nebula emission.
The images were then denoised with {\sc Neat Image},  V.8\footnote{https://ni.neatvideo.com/features/version-history/ni8ps} without  changing the structures in the nebula and  increasing the contrast with respect to the background. 
Finally, the images were stacked with {\sc Photoshop}, and the stars re-added to the final frame. The final image is shown in Fig.~\ref{fig:2}. 
The individual night images in the \OIII\, filter were treated in an identical way, producing a final image shown in  Fig.~\ref{fig:3}. For the fewer short-exposure images in the  \SII\, filter we adopted a simpler procedure, stacking them and enhancing  the contrast  (Fig.~\ref{fig:4}).

\subsection{Spectroscopic data reduction}
Data reduction of the long-slit spectra was performed in a standard way as described in  previous papers \citep[for example][]{Egorov2018}. The spectra of spectrophotometric standard stars obtained during the same nights were used to  calibrate the long-slit spectra to the absolute flux.

\section{Data analysis}
\label{sec:analysis}
\begin{figure*}
   \centering
   \includegraphics[width=\hsize]{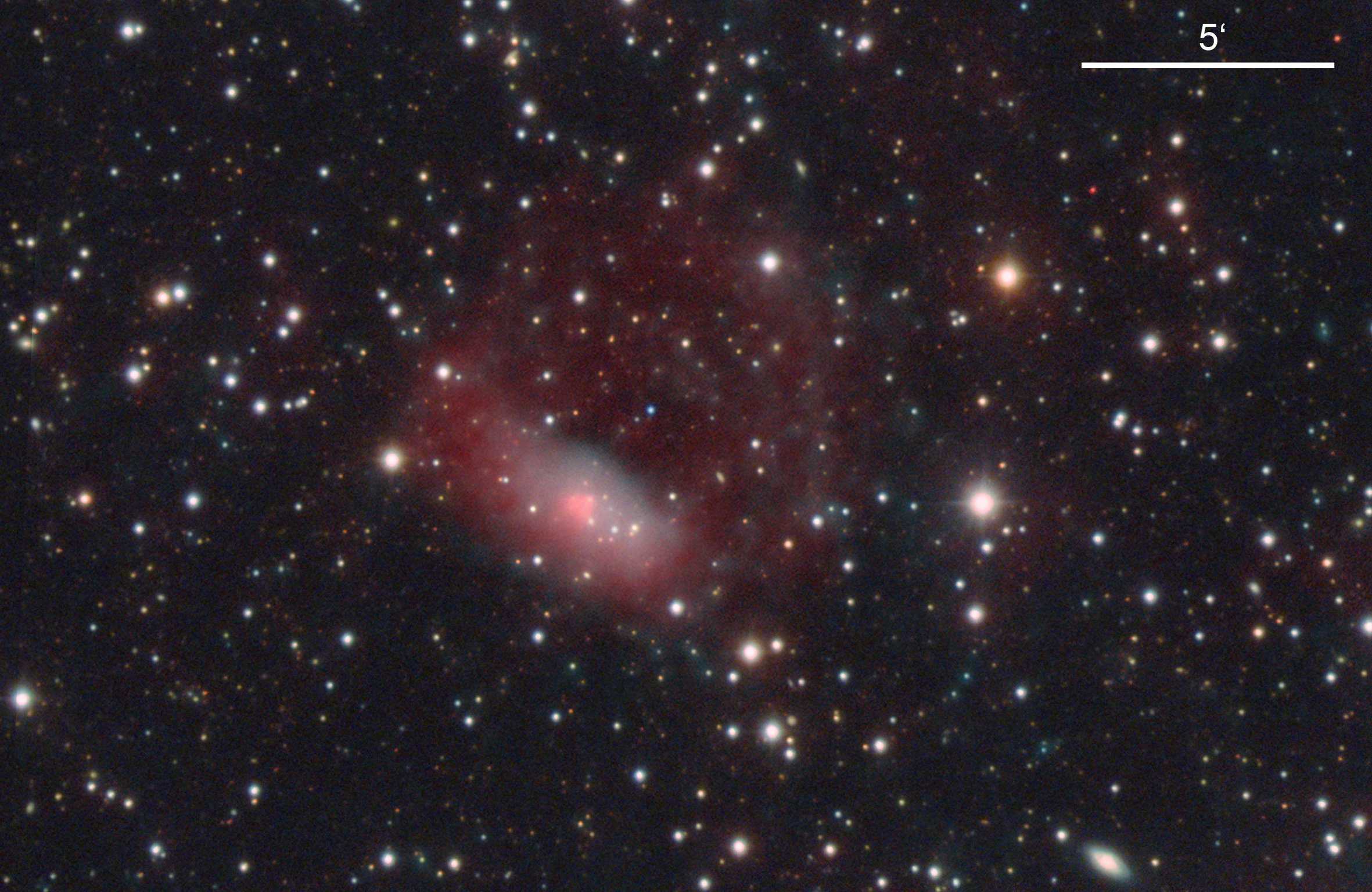}
      \caption{Colour composite image of TBG-1: H$\alpha$+\NII{} filter as in Fig.~\ref{fig:3} (red channel),  \OIII\, filter as in   Fig.~\ref{fig:4} in the  blue and green channels, and near infrared. 
      North is up, east to the left.
      }
         \label{fig:5}
   \end{figure*}

\begin{figure}
   \centering
   \includegraphics[width=\hsize]{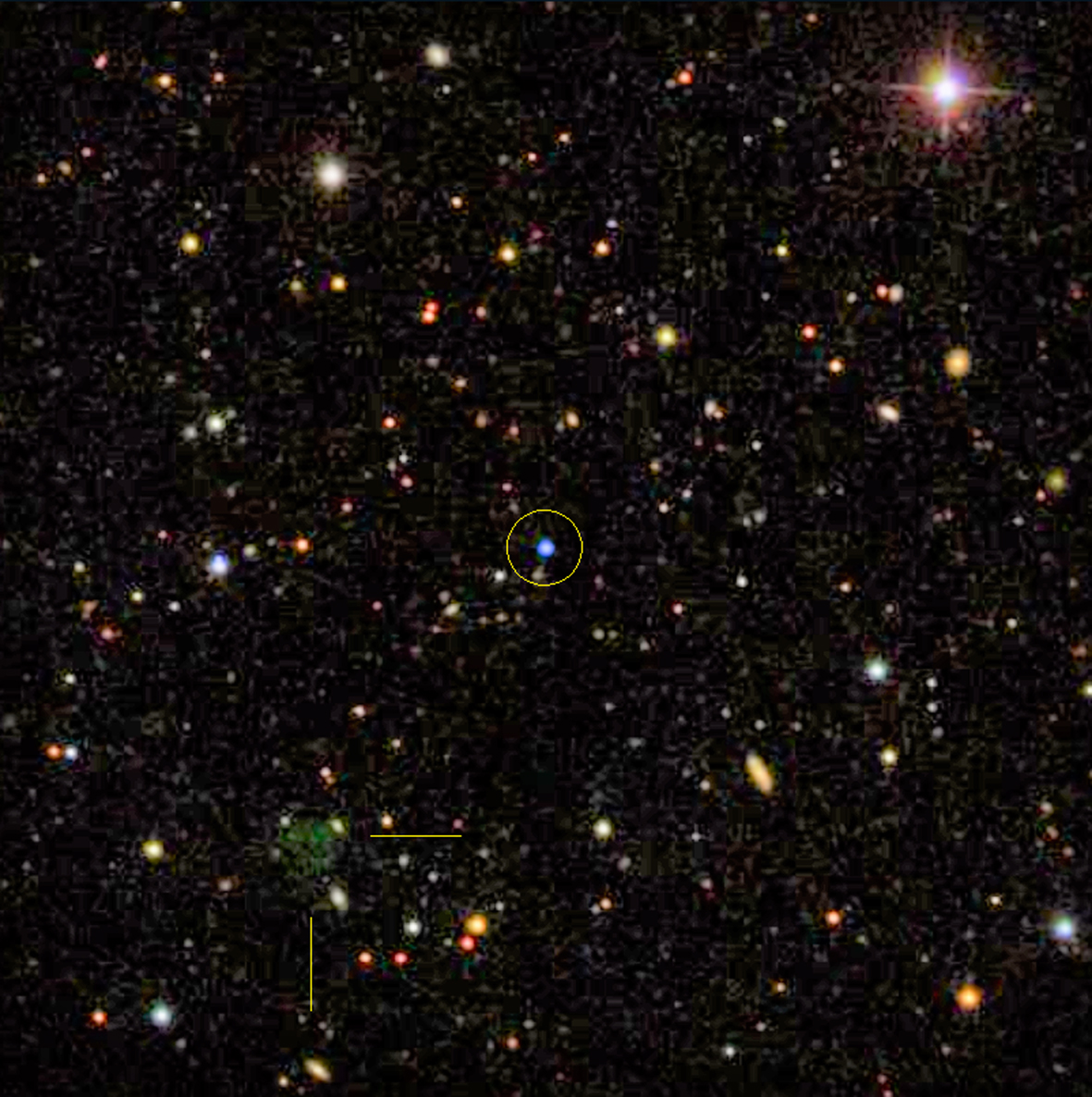}
      \caption{The Triangle nebula (marked with two yellow lines) and the  central star SDSS J075820.03+664558.6 (circle) in  the SDSS DR16 image. The bright star in the upper right corner is 3UC314-
042886 with a visual magnitude of 12.85. North is up, east to the left.
              }
         \label{fig:6}
   \end{figure}

\begin{figure*}
   \centering
   \includegraphics[width=\hsize]{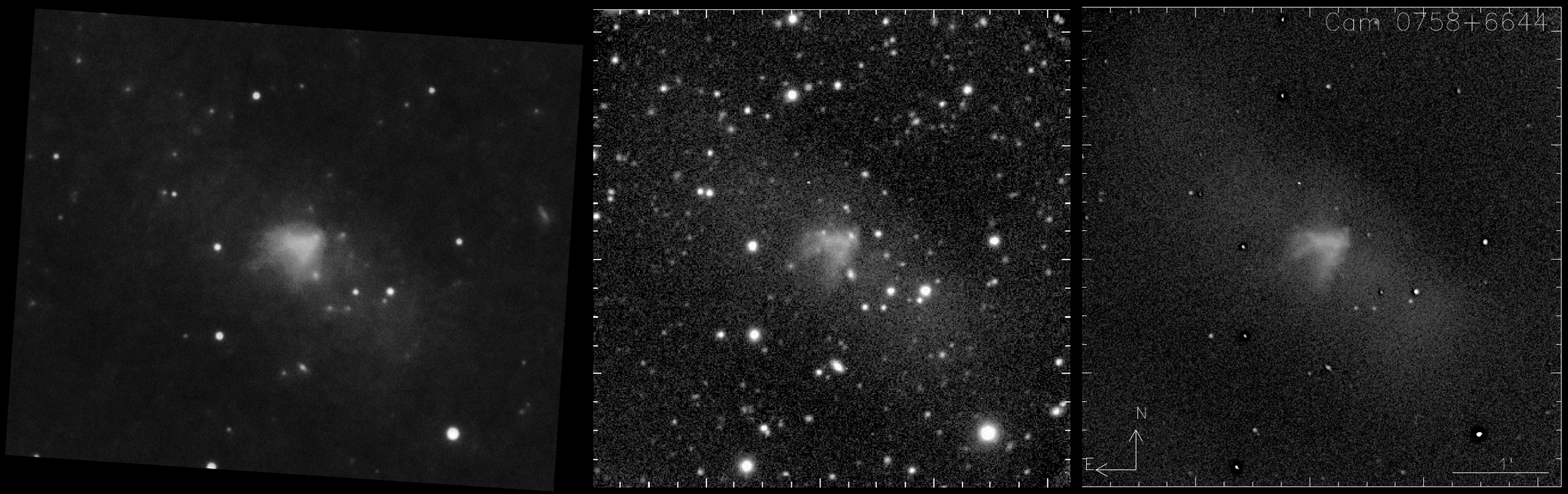}
  \caption{Sub-structures within the Cam nebula in the neighbourhood of the Triangle nebula. Left panel: long exposure with the amateur TBG telescopes. Centre and left panels: observations with the SCORPIO-2/BTA original and continuum subtracted  H$\alpha$+\NII{} images. All images are with the same scale and orientation.
              }
         \label{fig:7}
   \end{figure*}

The visual inspection of Fig.~\ref{fig:5}  gives us important information about the structure of the nebula: {\em i)} in the brightest parts of the nebula, the triangular structure (the “Triangle Nebula”) appears red, i.e. it is  weaker in \OIII\, than in H$\alpha$+\NII. {\em ii)} The Cam nebula located around the triangular one is as bright in \OIII\, as in H$\alpha$+\NII. It is clearly shifted to the northwest, i.e. towards the centre of the "shell". {\em iii)}  Close to the geometric center of the circular shell of TBG-1  there is a blue star, identified as SDSS J075820.03+664558.6  with  $V=17.4$ mag   (see Fig.~\ref{fig:6}). We discuss the nature of the central star in Section~\ref{sec:centralstar}. 
In Fig.~\ref{fig:7}, we highlight in the long-exposure luminance image and in H$\alpha$+\NII\, some substructures of the nebula: a parabolic-shaped structure around the Triangle nebula,  with very little contrast with respect to the Cam Nebula. The axes of symmetry of both the parabolic structure and the Triangle nebula appear to be identical. The vertex of the parabola is pointing approximately to the central star position.
Directly adjacent to the vertex and in the axis of symmetry of the parabolic and triangular structure, there is a darker region in the Cam nebula, less extended than the Triangle nebula. The continuum-subtracted H$\alpha$+\NII{} image of the Triangle nebula obtained with SCORPIO-2/BTA in Fig.~\ref{fig:7} confirms the emission-line nature of the nebula. 

\begin{figure}
   \centering
   \includegraphics[width=\hsize]{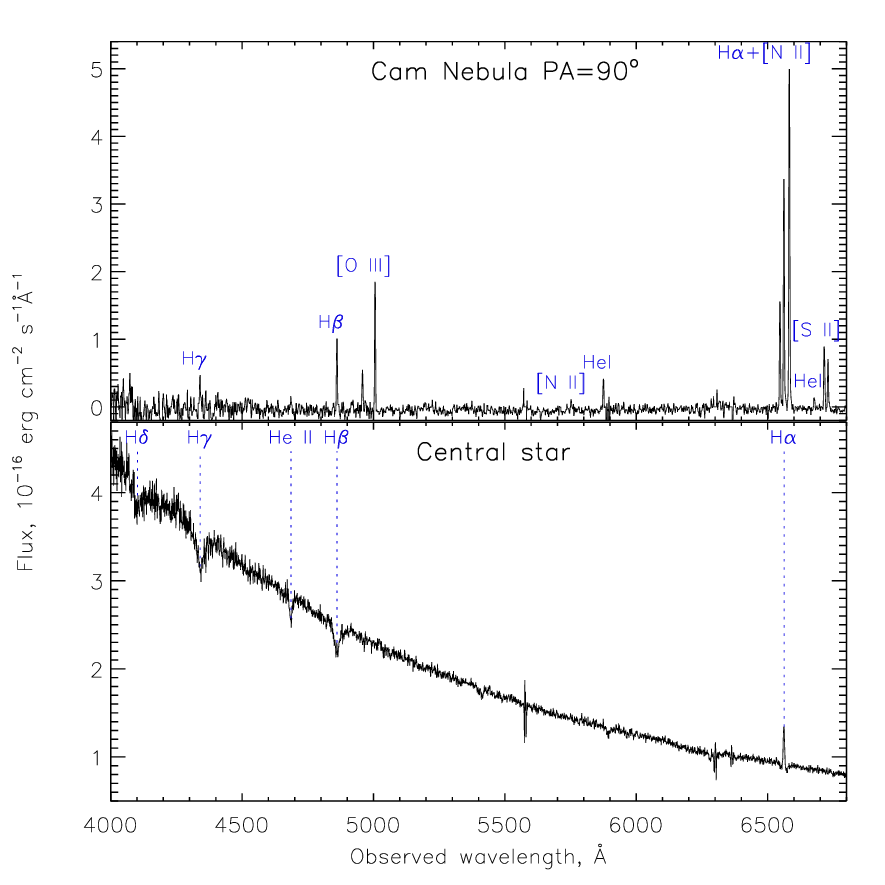}
      \caption{The integrated SCORPIO-2 spectrum of the brightest part of nebula (top) and of the central star (bottom). The main emission and absorption lines are labelled.
              }
         \label{fig:spectra}
   \end{figure}

Prompted by the confirmation of the nebula's nature as an emission-line object, we obtained both the nebular spectrum and that of the central star. The spectra are  shown in Fig.~\ref{fig:spectra}. 
The spectrum obtained with a slit along $PA=89\degr$ 
reveals numerous emission lines produced by the ionised gas: the hydrogen Balmer lines, the collisional excited lines of \OIII, \NII\, and \SII, and  a faint but detectable emission line due to  \HeI\, (see Fig.~\ref{fig:spectra}). The spectrum with $PA=3\degr$ across the shell does not show emission-line features since the emission of the shell is too weak, but it clearly shows the absorption spectrum of the central star.   
The only exception is the emission line of \Ha, which overlaps with the absorption line present in the stellar spectrum. This emission may be produced by the portion of the nebula that is intercepted by the slit located at $PA=3\degr$, or, as discussed below, can be produced by the central star itself.  

\begin{figure}
   \centering
   \includegraphics[width=\hsize]{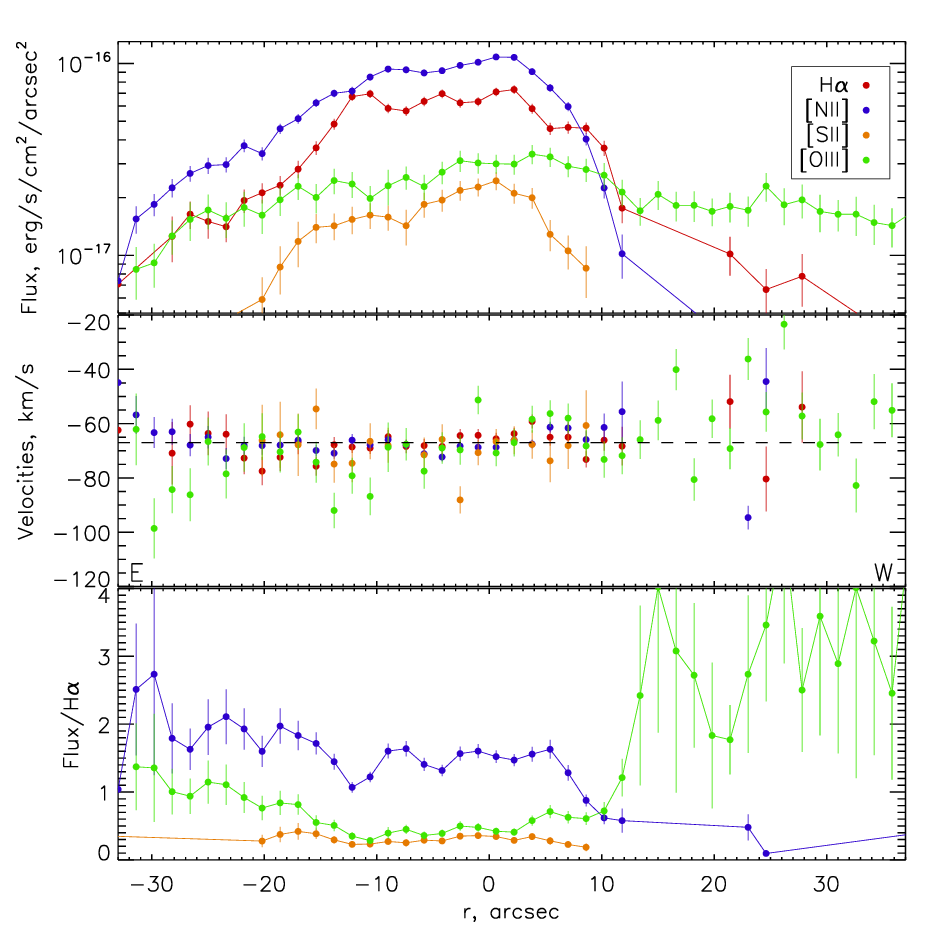}
      \caption{Radial variations along the slit of the flux of each emission line, of radial velocities  computed with each line, and of flux ratio relative to the \Ha. The error bars correspond to $3\sigma$ level. The centre of the scale correspond to the triangle nebula, negative values to the east, and positive to the west.  }
         \label{fig:lines}
   \end{figure}

In Table~\ref{table:tab1}, we present  the integrated fluxes, normalised to H$\beta$=100,  of the detected emission lines. The fluxes are  obtained with a  single-component Gaussian fitting and are  given separately for the brightest part of the nebula ($\sim45\arcsec$ in size along the slit, "The core'') and for the larger  region ($\sim110\arcsec$, "total'') which is  an area with a smoother distribution of the line emission dominated by the \OIII\, emission. In Fig.~\ref{fig:lines}, we show the radial variation of the flux of the main emission lines, together with the variation of the radial velocity measured from different emission-lines, and the flux ratio with respect to H$\alpha$.  The \OIII/\Ha{} flux ratio increases in the western side of the slit. 

In Table~\ref{table:tab1}  we show  some physical quantities derived from the emission-line  ratios in the both integrated spectra: dust extinction A$_V$ obtained from the comparison between the  observed H$\alpha$/H$\beta$  ratio and the intrinsic ratio for Case~B recombination \citep{Osterbrock2006agna.book.....O} and the reddening law of \citet{Cardelli1989ApJ...345..245C}; the electron temperature  $T_e$ from the  \NII ($\lambda6548$ + $\lambda6584$)/$\lambda5755$ ratio, using the  calibration  for the low density regime \citep{Pilyugin2010ApJ...720.1738P}; and electron density $n_e$ estimated from sulphur doublet flux ratio \SII$\lambda6716$/$\lambda6731$ with the calibration  of \citet{Proxauf_2014}. 
Unfortunately, the observed \SII$\lambda6716$/$\lambda6731$ corresponds to the degeneracy regime between this ratio and $n_e$. The uncertainties are very large, showing that we are in the degeneration regime, and confirming that we can give only upper limits.    Strictly speaking, we can only conclude, that $n_e<50-80$ cm$^{-3}$. In Table~\ref{table:tab1}, we show the formal values obtained for the electron density with their uncertainties.  

\begin{figure}
   \centering
   \includegraphics[width=\hsize]{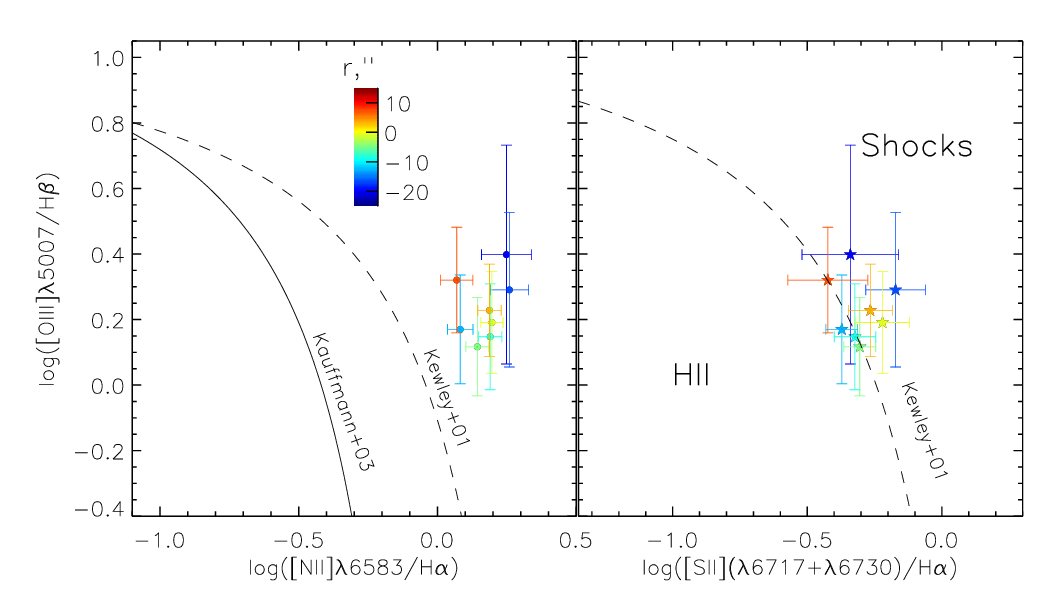}
   \includegraphics[width=\hsize]{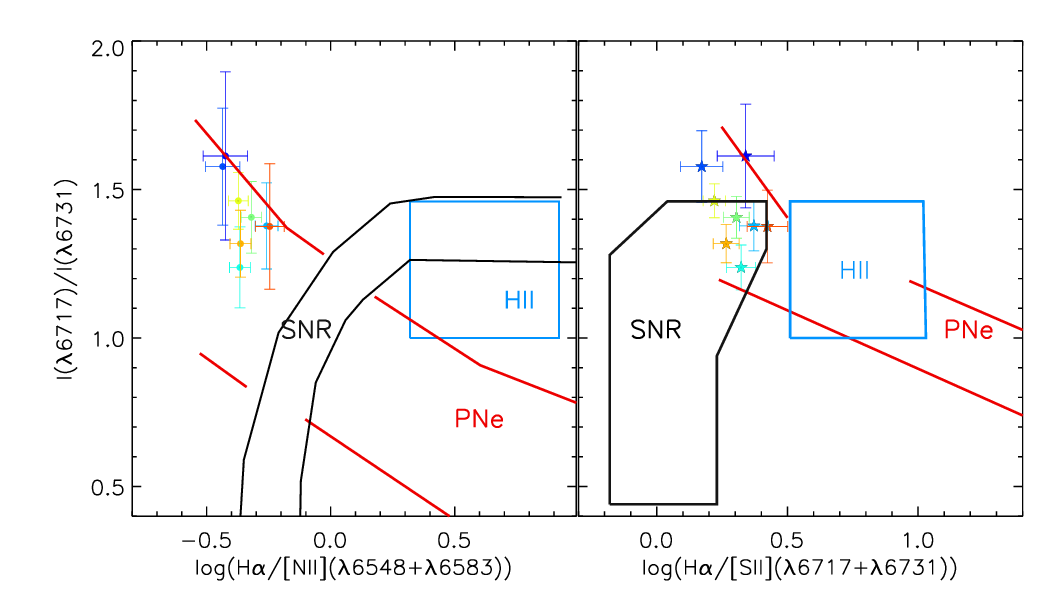}
      \caption{The diagnostic emission line BPT  diagrams (top panel). The  dividing lines between the regions ionised by hot stars and shocks are shown according to \citet{Kewley2001ApJ...556..121K,Kauffmann2003MNRAS.346.1055K}. The bottom panels show diagrams used for separation between SNR, HII regions, and PNe   \citep[dividing lines  were adopted from][]{Leonidaki2013MNRAS.429..189L}. The colour box corresponds to different locations along the slit for the brightest area of the nebula. 
        }
         \label{fig:ratio}
   \end{figure}

From the trend of radial velocity as a function of the position along the slit, we can deduce that the gas is dynamically cold. We do not appreciate any detectable velocity gradient along the slit: the mean heliocentric velocity in all lines is $V(hel)=67\pm8\kms$ (Fig.~\ref{fig:lines}). The lines profile has a single-component structure without broadening (the observed $FWHM\approx200\kms$ corresponds to the instrumental resolution). 
In Fig.~\ref{fig:ratio}, we show the  diagnostic diagrams based on ratio of pairs of lines adjacent  in wavelength, thus independent of interstellar extinction: BPT diagrams  \citep*[after][]{Baldwin1981PASP...93....5B} and  diagrams used to separate PNe, H~{\sc ii} regions and  supernova remnants (SNRs) as in \citet{Leonidaki2013MNRAS.429..189L}. It can be clearly seen in both diagrams  that the line ratios  correspond to emission from  PNe or SNRs.

\begin{table}
\caption{Relative intensities of emission lines in the brighter part of the
Can Nebula [I(\Hb) = 100, 1$\sigma$ errors] and the ionised gas parameters.}             
\label{table:tab1}      
\centering                          
\begin{tabular}{lrr}        
\hline\hline                 
Lines &  The core & total \\    
\hline                        
H$\gamma\lambda$4340 &    41.0$\pm$3.3 &     42.2$\pm$ 3.2 \\ 
 H$\beta\lambda$4861 &   100.0$\pm$1.9 &    100.0$\pm$ 2.4\\ 
  \OIII$\lambda$5007 &   179.9$\pm$4.0 &    263.0$\pm$ 6.7\\ 
   \NII$\lambda$5755 &    15.5$\pm$1.0 &     16.8$\pm$ 1.1\\ 
   \HeI$\lambda$5876 &    43.7$\pm$1.8 &     41.0$\pm$ 2.1\\ 
   \NII$\lambda$6548 &   150.5$\pm$3.2 &    125.7$\pm$ 3.5\\ 
H$\alpha\lambda$6563 &   325.8$\pm$6.6 &    310.4$\pm$ 7.6\\ 
   \NII$\lambda$6583 &   480.0$\pm$9.5 &    419.9$\pm$10.1\\ 
   \HeI$\lambda$6678 &    17.3$\pm$1.6 &     17.7$\pm$ 1.7\\ 
   \SII$\lambda$6716 &    93.1$\pm$2.2 &     81.7$\pm$ 2.5\\ 
   \SII$\lambda$6731 &    66.6$\pm$1.7 &     57.0$\pm$ 1.9\\ 
\hline        
A$_V$, mag         & 0.41$\pm$0.05 & 0.28$\pm$0.20 \\
 $T_e$(\NII), K          & 13\,700$\pm$350 & 14\,800$\pm$1200 \\
$n_e$, cm$^{-3}$   & 28$\pm$ 54     &  2$\pm$54   \\
\hline
\end{tabular}
\end{table}

\section{The properties of the nebula and of the  central star}
\label{sec:centralstar}

  \begin{figure}
   \centering
   \includegraphics[width=\hsize]{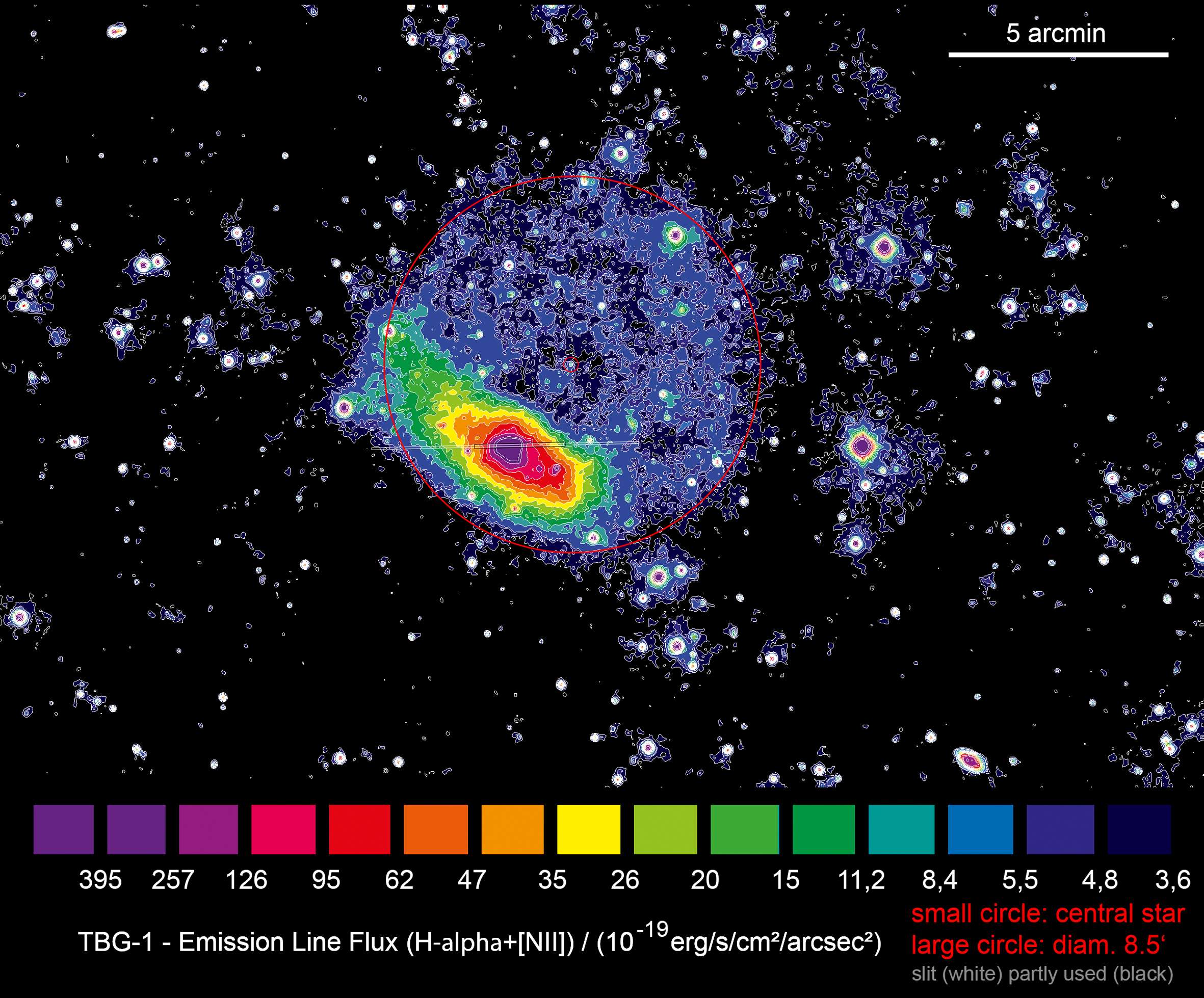}
      \caption{Flux calibrated image of TBG-1 and the surrounding star field in the \Ha +\NII\, filter. The flux scale is included. The  estimated circumference of the shell (large red circle) with a radius of 8\arcmin.25 around the central star is shown, and the central star is marked with a small red circle. The position of the slit with PA = 89$^{\circ}$ is indicated. }
         \label{fig:13}
   \end{figure}

We flux calibrated  the deep image in the \Ha + \NII{} filter (Fig.~\ref{fig:2}) by using the measurement of intensity variations in these lines along the slit at PA=89$^{\circ}$. In Fig.~\ref{fig:13} we show the flux-calibrated image of TBG-1. From this image, we can estimate  the diameter of the shell nebula d=8\arcmin.5$\pm$0\arcmin.5. Assuming the distance of the shell to be the same as the central star candidate one (discussed in detail in this section), D=1,019$\pm$89 pc, the angular size corresponds to a linear shell radius of 1.26$\pm$0.15 pc.

Assuming the electron density scales with the surface brightness of the nebula, we might estimate an upper limit from  n$_{e}<$2 to n$_{e}<$9 cm$^{-3}$ for the electron density of the shell (fainter and brighter regions, respectively). 
This estimate is obtained scaling our most conservative estimate of the electron density , i.e. the upper limit n$_{}e<$80 cm$^{-3}$, and scaling that value with the ratio between the flux in the Cam nebula and the shell. Modelling the shell with a Str\"omgren sphere, we would need an effective electron density not larger than 0.15~cm$^{-3}$. This condition is fulfilled assuming a n$_{e}<2$~cm$^{-3}$ and a filling factor within the shell of 0.07 or smaller.

Assuming that the distribution of the ionised gas within the nebula corresponds to the relative line intensity ratios (Table~\ref{table:tab1}) and using a mean electron density of $\sim$2~cm$^{-3}$ and a filling factor of 0.07,  we estimate the total mass of the shell M$_{PN}$$\sim$0.38~M$_{\odot}$. Both values (filling factor and total mass) are in agreement with the results of \citet{mallik1988RMxAA..16..111M}, who found that the larger the radius of the planetary nebula, the smaller the filling factor, and that there is   a linear relationship between mass and radius of planetary nebulae.
Considering a typical  expansion velocity of the shell  between 20 and 40 km~s$^{-1}$,  the time needed to reach that radius is between 30,000 and 60,000 yr. This is a very long time, which might make the discovered PN one of  largest planetary nebula within the Galaxy and probably one of the oldest.

The star located at the centre of the spherical nebula, SDSS J075820.02+664558.9, also identified in  {\em Gaia} {\sc DR3}  1095335102795586944, and classified  as a White Dwarf (WD) candidate (V=17.4  and A$_V$=0.11) might be the central star (CS) of the nebula. 
Using the distance of 1.019 kpc derived from the inverse of parallax according to {\em Gaia} {\sc DR3} catalogue, the absolute V magnitude is M$_V$=7.3. 
To understand if this star might be able to excite the nebula, we calculate its Str\"omgren sphere radius. 
The {\em Gaia} database \citep{gaia2023A&A...674A...1G} gives an effective temperature T$_{\rm eff}$=22\,800$\pm$1,000 K, which is in agreement with the value obtained by fitting a black body to the SCORPIO-2 spectrum,  $_{\rm eff}$=21,000-24,000 K. 
{\em Gaia} {\sc dr3} provides also the mass of H and He. Using the relationship between mass and radius for WD \citep[see, e.g.][]{Koester1990RPPh...53..837K, Joyce2017ASPC..509..389J, Romero2019MNRAS.484.2711R}, we obtain R$_{\rm CS}>$0.015 R$_\odot$. 
Using the electron density and temperature estimated from the nebular spectrum, and the properties of the central star,  we obtain a  radius of the Str\"omgren sphere in agreement with the observed radius of the shell. Therefore, the properties of the star are consistent with those of a PN central star. 
From Fig.~\ref{fig:spectra}, we notice that the spectrum of the CS shows a component in emission for both  \Ha\, and  \Hb\, lines. The emission could be due to the overlap between the stellar and nebular spectra. However, the surface brightness of TBG-1 in \Ha+\NII\, at those coordinates is likely too low to produce the observed effect. Therefore, it should be a  peculiar characteristics of the CS, which are accompanied, for example, by its position in the HR diagram somewhat outside the main locus of white dwarfs   (at the top, slightly shifted to the right, with respect to WD locus in the {\em Gaia} HR diagram of WDs\footnote{https://sci.esa.int/web/gaia/-/60209-white-dwarfs-in-gaia-s-hertzsprung-russell-diagram}). A more detailed study of the PN and of its CS is needed. In particular, we would need deeper images  in the \OIII\, and \SII\, narrow-band filters to improve the signal-to-noise ratio and determine the line ratios in the weaker parts of the shell of TBG-1. We also aim at acquiring  high-resolution spectra of the CS to determine its radial velocity in order to verify the agreement between the motion of the PN and the CS. High-resolution spectra of the CS would be used to investigate the nature of the emission in \Ha\, and in \Hb.

\section{Discussion and Conclusions}
\label{sec:conc}
In this paper we report the discovery and characterisation of a 
new planetary nebula, named TBG-1. The nebula is composed of three structures:
the "Triangle nebula” is the brightest part of the nebula complex and it shows filamentary structures, arranged in a triangular shape. It was detected in \Ha, \NII, \SII\, and \OIII;  a faint parabolic structure around the “Triangle nebula”  observed in \Ha+\NII,  possibly part of a bipolar structure, the second half of which is hidden behind a local dust cloud (see Fig.~\ref{fig:7});  an elliptical nebula structure, named Cam nebula surrounds the Triangle nebula; and a circular faint shell, with a blue star at its centre.

Because  the central star is located almost exactly at the centre of the circular nebula, a physical connection between the two objects  might be expected. Therefore we assume that it is the central star, and that the distance to the central star is the distance to the nebula. Accordingly to this distance estimate,  the nebula might be one of the largest and oldest planetary nebulae ever discovered. This is also supported by the extremely low surface brightness of the object. 
The central star, with an estimated effective temperature of around 22\,800 K, is capable of ionising a Str\"omgren sphere with a radius close to  the observed one. 
This work shows how amateur studies can lead to excellent scientific results when done with passion and scientific rigour. Large surveys are increasing the discovery space for many objects, but peculiar objects, such as this PN, which is extremely faint, can only be detected with long exposures and thorough analysis.

\begin{acknowledgements}
      Part of this work was supported by the German
      \emph{Deut\-sche For\-schungs\-ge\-mein\-schaft, DFG\/} project
      number Ts~17/2--1.
      LM acknowledges INAF for the Minigrant 2022 CHECS and for the Large Grant 2023 EPOCH. LM acknowledges the grant PRIN project n.2022X4TM3H  "Cosmic POT" from Ministero dell'Universit\'a e la Ricerca (MUR). We obtained part of the observed data on the unique scientific facility `Big Telescope Alt-azimuthal' of SAO RAS as well as analysed the spectral and imaging data  with the financial support of grant No~075-15-2022-262 (13.MNPMU.21.0003) of the Ministry of Science and Higher Education of the Russian Federation.
      Funding for the Sloan Digital Sky Survey V has been provided by the Alfred P. Sloan Foundation, the Heising-Simons Foundation, the National Science Foundation, and the Participating Institutions. SDSS acknowledges support and resources from the Center for High-Performance Computing at the University of Utah. The SDSS web site is \url{www.sdss.org}.

\end{acknowledgements}

%
 \bibliographystyle{aa} 
\bibliography{Ref_spec} 

\end{document}